\documentclass[10pt,a4paper,twocolumn,superscriptaddress]{revtex4-1}
\usepackage[latin1]{inputenc}
\usepackage{amsmath}
\usepackage{amsfonts}
\usepackage{amssymb}
\usepackage{graphicx}
\usepackage{color}
\usepackage{float}
\usepackage{multirow}
\usepackage{booktabs} 

\newcommand{\ddl}{\mathrm{d}}
\newcommand{\vect}{\mathbf}
\usepackage[range-units=single,separate-uncertainty]{siunitx}
\begin{document}
\author{R. St\"uhler}
\author{F. Reis}
\affiliation{Physikalisches Institut and R\"ontgen Research Center for Complex Material Systems, Universit\"at W\"urzburg, D-97074 W\"urzburg, Germany}
\author{T. M\"uller}
\author{T. Helbig}
\author{T. Schwemmer}
\author{R. Thomale}
\affiliation{Institut f\"ur Theoretische Physik und Astrophysik, Universit\"at W\"urzburg, D-97074 W\"urzburg, Germany}
\author{J. Sch\"afer}
\email{e-mail: joerg.schaefer@physik.uni-wuerzburg.de}
\author{R. Claessen}
\affiliation{Physikalisches Institut and R\"ontgen Research Center for Complex Material Systems, Universit\"at W\"urzburg, D-97074 W\"urzburg, Germany}
\date{\today}


\title{ 
Tomonaga-Luttinger liquid in the edge channels of a quantum spin Hall insulator
}
\maketitle


\textbf{
Topological quantum matter is characterized by non-trivial 
global invariants of the bulk which induce gapless electronic 
states at its boundaries. A case in point are two-dimensional 
topological insulators (2D-TI) which host one-dimensional (1D) 
conducting helical edge states protected by time-reversal symmetry 
(TRS) against single-particle backscattering (SPB). However, as
\textit{two}-particle scattering is not forbidden by TRS~\cite{Xu2006}, the
existence of electronic interactions at the edge and their notoriously 
strong impact on 1D states may lead to an intriguing interplay between 
topology and electronic correlations. In particular, it is directly relevant 
to the question in which parameter regime the quantum spin Hall effect (QSHE) 
expected for 2D-TIs becomes obscured by these correlation effects that prevail 
at low temperatures \cite{Li2015}. Here we study the problem on bismuthene on 
SiC(0001) which has recently been synthesized and proposed to 
be a candidate material for a room-temperature QSHE~\cite{Reis2017}. 
By utilizing the accessibility of this monolayer-substrate system on atomic
length scales by scanning tunneling microscopy/spectroscopy
(STM/STS) we observe metallic edge channels which display 1D electronic correlation effects. 
Specifically, we prove the correspondence with a Tomonaga-Luttinger 
liquid (TLL), and, based on the observed universal scaling of the differential
tunneling conductivity ($dI/dV$), we derive a TLL parameter $K$ reflecting
intermediate electronic interaction strength in the edge states of bismuthene. 
This establishes the first spectroscopic identification of 1D electronic 
correlation effects in the topological edge states of a 2D-TI. 
}


The topological protection of the 1D metallic edge channels in 2D-TIs against 
elastic SPB by TRS~\cite{Kane2005, Kane2005a} leads
to quantized, i.e. dissipationless transport which is reflected in the QSHE.
Moreover, the property of spin-momentum locking renders 2D-TIs 
promising candidate materials for applications in spintronics. To date, 
the QSHE has only been measured in three material systems that are all 
characterized by small bandgaps ($E_{\text{gap}} \leq \SI{55}
{\milli\electronvolt}$) of which the quantum well (QW) structures of 
three-dimensional semiconductors, such as HgTe/CdTe~\cite{Koenig2007}
and InAs/GaSb~\cite{Knez2011}, constitute the most prominent realizations. 
Recently, the QSHE effect has been reported to be observed up 100 K in 
monolayer crystals of WTe$_2$ \cite{Wu2018,Tang2017}. While most 
of these experiments can be well-understood within topological band
theory of non-interacting electrons, the deviations from a sharply quantized
conductance ($2e^2/h$) seen at very low temperatures in InAs/GaSb QW transport measurements
have been attributed to the relevance of electronic interactions \cite{Li2015}. In fact, in any 
real 2D-TI such interactions are inevitably present and will play a 
non-negligible role for the 1D edge states which are consequently 
expected to constitute a helical TLL.~\cite{Wu2006,Xu2006} A direct spectroscopic 
identification of this correlated many-body state 
in the edge states of a 2D-TI, however, has remained lacking so far.   


Bismuthene, i.e.~a 2D monolayer of Bi atoms epitaxially grown on
a semiconducting SiC(0001) substrate, turns out to be a particularly 
well-suited system for such studies. Here the strong atomic 
spin-orbit coupling in the Bi atoms conspires with their honeycomb 
arrangement and covalent coupling to the substrate to drive the 
system into a non-trivial topology \cite{Reis2017,Li2018}.
Angle-resolved photoelectron spectroscopy as well as STM/STS 
found excellent agreement with the calculated non-trivial 
topological band structure \cite{Reis2017} and confirmed 
a sizable fundamental band gap 
($\Delta \sim \SI{0.8}{ \electronvolt}$), potentially allowing 
the persistence of the QSHE up to room-temperature and
beyond. Moreover, STM/STS measurements clearly indicate 
the existence of conducting edge channels which due to the 
large bulk band gap display an extremely small exponential 
decay length $\xi$ into the insulating 2D bulk of only 
$\SI{4.1}{\angstrom}$ \cite{Reis2017}.


\begin{figure*}
	\centering
	\includegraphics[width = 0.95\textwidth ]{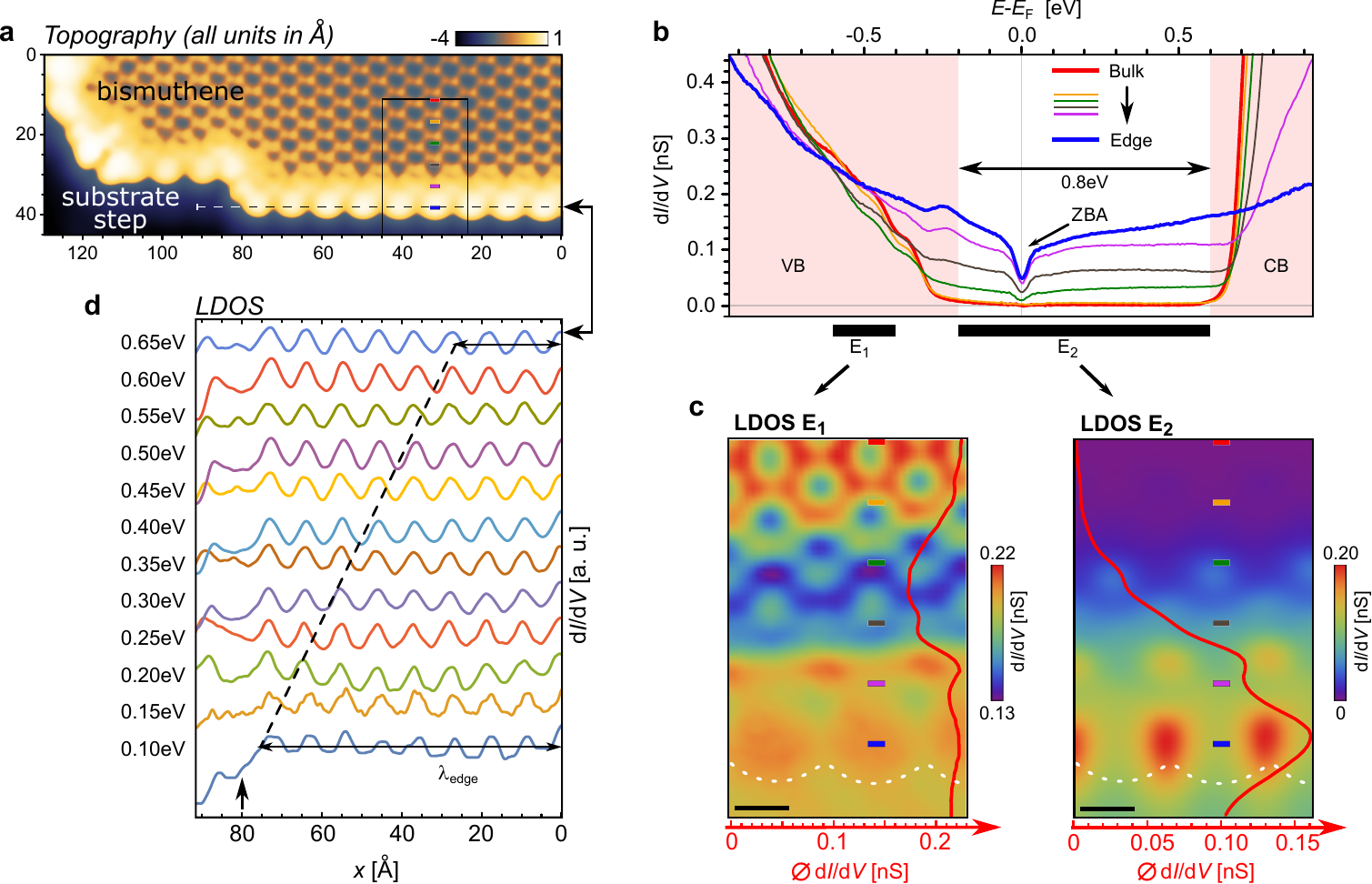}
	\caption{ 
		\textbf{Strongly confined 1D metallic channels at bismuthene armchair edges.}
		\textbf{a}, Constant current image (topography) of a bismuthene armchair edge terminated 
		by a SiC terrace step ($T = \SI{4.4}{\kelvin}$, 
		$V_{\text{set}} = \SI{-1.1}{\volt}$, $I_{\text{set}} = \SI{100}{\pico
			\ampere}$).
		\textbf{b}, $dI/dV(V)$ spectra at positions indicated by the colored marks 
		in a and c. The edge LDOS is suppressed $\pm \SI{100}{\milli \electronvolt}$ 
		around $E_\text{F}$ (ZBA).
		\textbf{c}, (left) Spatially resolved LDOS-map averaged over an energy range $E_1: \SI{-0.6}{\electronvolt} \leq E-E_\text{F} 
		\leq \SI{-0.4}{\electronvolt}$ in the occupied 
		states. The bismuthene bulk LDOS can clearly be seen 
		as honeycombs. (right) Spatially resolved LDOS map averaged over the bulk 
		energy gap $ E_2: \SI{-0.2}{\electronvolt} \leq E-E_\text{F} \leq \SI{0.6}
		{\electronvolt}$. The bulk states disappear and only a $\sim \SI{1.5}
		{\nano \meter}$ wide metallic channel remains at the edge that extents 
		along the terrace step. White dashed lines: Edge in the topography. Red curves: Spatially and energetically 
		averaged differential conductivities perpendicular to the edge. Within 
		the bandgap the averaged differential conductivity confirms the exponential 
		decay ($\xi = \SI{4.1}{\angstrom}$) of the edge LDOS towards the bulk. The scale bar corresponds to $\SI{5}{\angstrom}$. (See movie in supplementary information.)
		\textbf{d}, Waterfall plot of LDOS line profiles at different energies measured along 
		the dashed line in a. The double arrow marks the electron wavelength
		$\lambda_{\text{edge}}$ according to DFT calculations.~\cite{Reis2017} The black arrow indicates the position of the kink in the bismuthene edge along the x-direction.
	}
	\label{fig:1}
\end{figure*}


The near-to-perfect spatial 1D confinement of the metallic edge 
states is ideally matched to the atomic resolution of STM/STS, 
as illustrated Fig.~\ref{fig:1}. While with constant current STM 
we are able to explore the surface topography, 
STS allows us to spatially map the local electronic density of states 
(LDOS). Accordingly, the STM topography map in 
Fig.~\ref{fig:1}(a) demonstrates the
successful synthesis of bismuthene with its planar atomic honeycomb
structure. While bismuthene covers SiC smoothly within substrate 
terraces, the monolayer film is truncated at SiC zigzag terrace 
steps. Due to the $\sqrt{3}\text{-R}30^\circ$ superstructure 
of the Bi honeycomb lattice with respect to the substrate, this 
translates into well-ordered armchair edges of the bismuthene 
layer (see Fig.~\ref{fig:1}(a) and Fig.~S1). We 
note that the periodic spacing between the bright white, 
oval edge features in Fig.~\ref{fig:1}(a) is in agreement 
with theoretical expectations for armchair edges and 
indicates the accumulation of charge density at
the edge.


This assumption is supported by the STS LDOS data. 
While the differential conductivity measured well inside the bulk 
material (Fig.~\ref{fig:1}(b), red curve) is dominated by the 
bulk bismuthene band gap, it 
becomes continuously filled as the tunneling tip moves towards 
the film edge (thin curves in Fig.~\ref{fig:1}(b)). Eventually, 
directly at the edge (blue curve in Fig.\ref{fig:1}(b)) the 
differential conductivity indicates a nearly constant LDOS, 
except for the apparent dip around $E_F$ indicated by 'ZBA' 
and discussed further below. Spatially resolved 
$dI/dV(V)$-maps taken parallel to the 
bismuthene edge (Fig.~\ref{fig:1}(c)) further elucidate the 
1D character of the edge LDOS. Within the occupied states, 
e.g., in the energy range $E_1$, the bismuthene 
bulk LDOS can be seen clearly. Evidently the edge LDOS has already set in. 
In contrast, inside the bulk energy gap $E_2$ the LDOS is confined solely to the 
geometrical edge, defining a highly constricted 1D metallic 
channel spreading along the entire extent of the SiC terrace
step edge ($\sim \SI{10}{\nano \meter}$ -- $\SI{200}
{\nano \meter}$). (See movie in supplementary information for full evolution of the edge LDOS with energy.)


We notice that the edge in Fig.~\ref{fig:1}(a) encompasses a kink 
in the substrate step. For an ordinary 1D electron wave one would 
expect a partial reflection off the kink causing a quasi-particle 
interference with the incident wave. Such emerging Friedel 
oscillations would scale with the energy-dependent electron wavelength 
$\lambda_{\text{edge}}$ defined by the band dispersion. \cite{Reis2017}
Eventually, the electrons in an ordinary 1D conductor may even 
become localized depending on the strength of the potential scatterer 
\cite{Meyer2003}. However, in our case the periodicity seen in the edge LDOS 
is exclusively caused by the topographic, i.e.~atomic modulation of 
the armchair edge, irrespective of electron energy as can clearly be 
seen in Fig.~\ref{fig:1}(d).
Most importantly, no energy-dependent Friedel oscillations 
are observed, consistent with the topological protection of the 
helical edge states in a 2D-TI against SPB.
Consequently, the effective mean free path of the 
edge channels is extraordinarily large, as kinks and (non-magnetic) impurity 
potentials do not act as scattering centers \cite{Pauly2015}.


\begin{figure}
	\centering
	\includegraphics[width = 0.45\textwidth ]{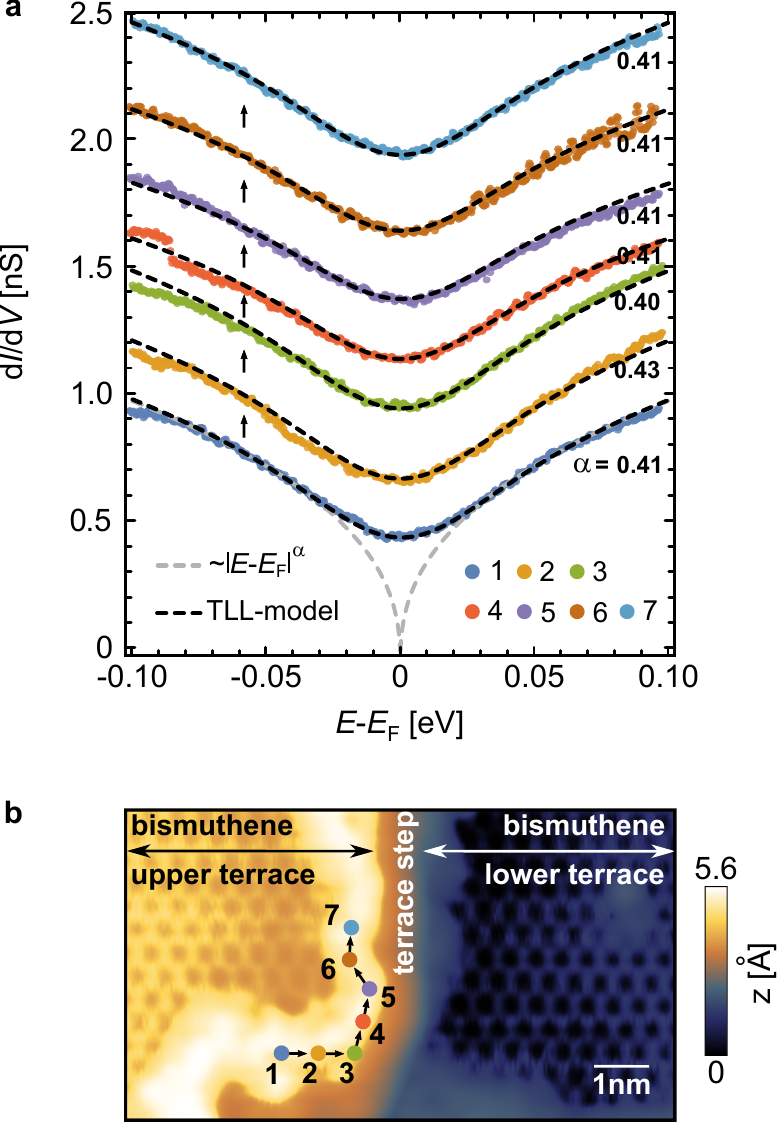}
	\caption{
\textbf{Power-law suppression of the ZBA.} 
\textbf{a}, Waterfall plot of single-point STS spectra along the path in 
b ($T = \SI{95}{\kelvin}$, 
$V_{\text{set}} = \SI{-0.1}{\volt}$, $I_{\text{set}} = \SI{300}{\pico\ampere}$). Spectra 2--7 are offset by multiples 
of \SI{0.25}{\nano\siemens}. The ZBA follows a power-law
$\sim |E-E_\text{F}|^\alpha$ (gray dotted line). The fitted 
power-law exponent $\alpha$ is listed on the right margin. 
Black dashed line: TLL model $\rho_{\text{TLL}}$ according to Eq.~\eqref{eq:TLL_LDOS} (taking into account thermal and instrumental broadening)
shows excellent agreement with the data over the entire 
energy range $\pm \SI{100}{\milli \electronvolt}$ around 
$E_\text{F}$. 
\textbf{b}, Constant current image of bismuthene at a 
SiC step ($T = \SI{95}{\kelvin}$, 
$V_{\text{set}} = \SI{-1.0}{\volt}$, $I_{\text{set}} = \SI{300}{\pico\ampere}$). STS spectra are marked with colored dots from 
$1-7$ along a path at the upper terrace that encounters a 
kink at the edge.
	}
	\label{fig:2}
\end{figure}


The LDOS of a (metallic) Fermi liquid is expected to be finite and 
nearly constant around the Fermi level. In contrast, our edge state 
spectra display a clearly detectable dip at $E_\text{F}$ which in 
the following will be referred to as zero-bias anomaly (ZBA). 
In an energy window of $\approx \pm \SI{100}{\milli \electronvolt}$ 
around $E_\text{F}$
it displays particle-hole symmetric behavior following (not too close 
to $E_\text{F}$) a power-law $\propto |E-E_\text{F}|^\alpha$ 
with $\alpha \sim 0.41$ as high-resolution single-point STS reveals
in Fig.~\ref{fig:2}(a). Interestingly, spectral shape 
and power law exponent $\alpha$ of the ZBA persist
along the entire length of the 1D edge channel and do not even
change as one follows the LDOS around a strongly kinky section 
of the edge (see Fig.~\ref{fig:2}(b) for measurement path),
reminiscent of the absence of SPB at the edge kinks.


\begin{figure*}
	\centering
	\includegraphics[width = 0.95\textwidth ]{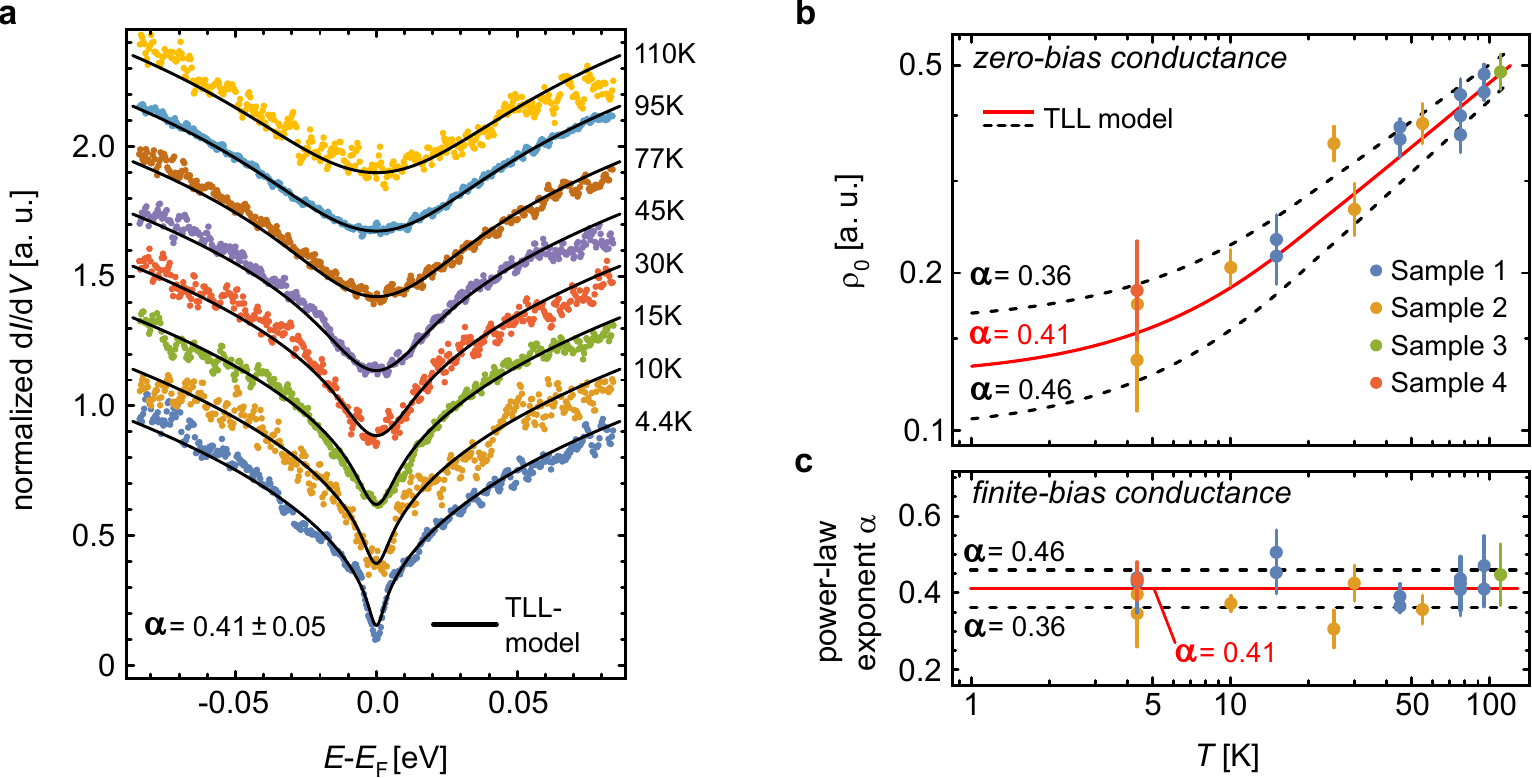}
	\caption{ 
	\textbf{Temperature dependence of the ZBA.} 
\textbf{a}, Waterfall plot of single-point STS spectra for different
temperatures. Spectra from \SI{10}{\kelvin}--\SI{110}{\kelvin} are offset by
multiples of 0.2. Towards higher temperatures the ZBA becomes continuously
filled. Black lines: all spectra are described by Eq.~\eqref{eq:TLL_LDOS} (taking into account thermal and instrumental broadening). The power-law exponent $\alpha$ is
the only fit parameter for the TLL model. 
\textbf{b}, Double-logarithmic plot of
the zero-bias conductance $\rho_0(T)$ as a function of temperature. Each data
point and error bar reflects the evaluation of $\sim 50$ spectra. Red and black
dashed lines: for $T > \SI{10}{\kelvin}$
$\rho_0(T)$ clearly follows a power-law $\rho_0 \sim T^{\alpha}$. Deviations from the bare power-law for $T < \SI{10}{\kelvin}$
are due to instrumental broadening. 
\textbf{c}, Power-law exponent $\alpha$ for
the finite-bias conductance $\rho(T)\sim |E-E_\text{F}|^\alpha $ as a function
of temperature. Each data point and error bar reflects the evaluation of $\sim 50$ spectra. Red and black
dashed lines: the energy power law remains constant 
as a function of temperature.
	}
	\label{fig:3}
\end{figure*}


The ZBA also features a pronounced temperature dependence, 
see Fig.~\ref{fig:3}(a). As the temperature increases 
from $\SI{4.4}{\kelvin}$ to $\SI{110}{\kelvin}$ the zero energy dip 
gets continuously filled. For further quantitative analysis 
Fig.~\ref{fig:3}(b) shows the normalized zero-bias differential conductivity
$dI/dV(V=0)$ as a function of temperature on a double logarithmic scale. It is
apparent that for $T > \SI{10}{\kelvin}$ the zero-bias differential conductivity
again follows a power-law $dI/dV(V=0)\sim T^{\alpha}$, with the same exponent 
$\alpha \sim 0.41$ already observed in the energy dependence.
For $T < \SI{10}{\kelvin}$ deviations from the bare power-law 
are due to instrumental broadening. The finite bias differential conductivity, 
on the other hand, and especially the energy power law remains constant 
as a function of temperature (see Fig.\ref{fig:3}(c)).


It is well-established that electronic interactions play a crucial role in 1D metallic
systems. As described by TLL theory, the Fermi liquid description breaks down in
1D and the physics is dominated by collective bosonic excitations~\cite{Voit1993,
Haldane1981}. Signatures of ordinary (spinful) TLLs have been observed, e.g., 
in semiconducting GaAs quantum wires \cite{Jompol2009}, metallic single 
wall carbon nanotubes \cite{Bockrath1999}, in polymer nanofibers 
\cite{Aleshin2004}, and atomic chains of gold atoms on Ge surfaces 
\cite{Blumenstein2011}. TLL behavior has also been identified in the 
chiral edge states of fractional quantum Hall systems \cite{Chang1996}.
A special case are the spin-momentum-locked edge states in a 2D-TI which 
form a \textit{helical} TLL as long as the electronic interaction is weaker 
than the band gap that protects them from the 2D bulk states 
\cite{Hohenadler2012}. However, one common characteristic inherent to all 
(spinful and helical) TLLs is a power-law behavior of their single-particle spectrum,
$\rho \sim |E-E_\text{F}|^\alpha$ for $|E-E_\text{F}| \gg k_B T$ 
\cite{Braunecker2012}, directly accessible by experimental electron 
removal and addition spectroscopies, such as STS \cite{Eggert2000}.
In fact, the energy dependence emerges from an even more fundamental
property of TLLs, namely the \textit{universal scaling} in both energy and 
temperature which is captured by the expression \cite{Bockrath1999}:
\begin{equation}
\rho_{\text{TLL}}(\epsilon,T) \propto T^\alpha \cosh\left( \frac{\epsilon}{2 k_B
T}\right) \Bigg|\Gamma\left(\frac{1+\alpha}{2} + i \frac{ \epsilon}{2 \pi k_B T}
\right) \Bigg|^2.
\label{eq:TLL_LDOS}
\end{equation}
This inherently generates also a temperature power-law dependence of the 
zero energy spectral weight with the same exponent as in energy.
Comparing with our experimental STS data we find excellent 
agreement with the theoretical $\rho_{\text{TLL}}$ over the
entire energy and temperature range as seen in Fig.~\ref{fig:2}(a) and 
Fig.~\ref{fig:3}(a)(black curves), suggesting that the ZBA observed in the 
edge states is indeed due to strong 1D electron corrrelations.


However, the TLL interpretation has to be discriminated against other 
possible origins of a ZBA. One scenario is the Efros-Shklovskii
pseudogap which can arise from the interplay of disorder and Coulomb 
interaction in low-dimensional metallic systems. It is characterized 
by an exponential suppression of the LDOS \cite{Bartosch2002}. 
As the observed ZBA in bismuthene follows a power-law instead, the 
disorder-induced pseudogap can be safely excluded (see also 
Fig.~S5 for a direct comparison).
Another mechanism that can lead to a suppression of
the low-energy tunneling is the dynamical 
Coulomb blockade (DCB) \cite{Hanna1991,Devoret1990} caused by 
the interplay between charge storage energy in and charge dissipation 
from a nanostructure (here the 1D edge). Within environment-quantum 
fluctuation or $P(E)$-theory this can be treated by modeling the
STS tunneling junction as a capacitance $C$ parallel to a resistance $R$.
For $|E-E_\text{F}| \ll E_c = e^2/2C$ and at zero temperature a power-law
behavior is predicted according to $dI/dV(E) \sim
|E-E_\text{F}|^{2/g} $ with the exponent $2/g = 2
R/R_K$ \cite{Devoret1990}, $R_K = h/e^2$ being the resistance quantum. 
We have measured the ZBA on four different samples, on
bismuthene domains with differing sizes, and with widely varied set-point parameters
$I_{\text{set}}$ and $V_{\text{set}}$ (see Fig.~S4), where 
the \textit{local} tunnel junction parameters $C$ and $R$ should widely differ accordingly. Yet, in all cases we observe the same
power law exponent $\alpha=0.41\pm0.05$, at variance with the DCB picture.
Finally, universal scaling would not be observed in the DCB scenario for the same locality reasons \cite{Ming2018}.

\begin{figure}
	\centering
	\includegraphics[width = 0.45\textwidth ]{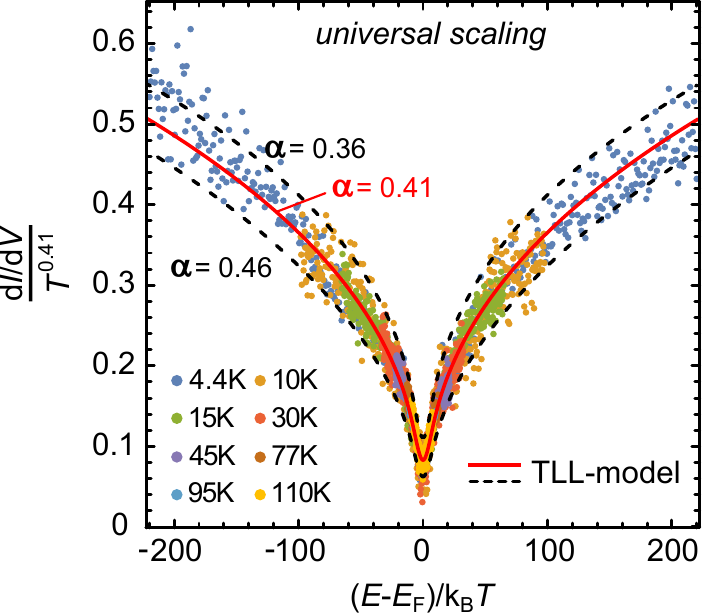}
	\caption{
\textbf{Universal scaling of the ZBA as a hallmark of a TLL.} On properly
rescaled axes the single-point STS spectra of Fig.~\ref{fig:3}(a) collapse
onto a single universal curve predicted as a hallmark for tunneling into a TLL.
Red curve: $\rho_{\text{TLL}}$ according to
Eq.~\eqref{eq:TLL_LDOS} (taking into account thermal and instrumental broadening) with exponent $\alpha = 0.41$. Black dashed curve: for
$\pm 0.05$ variation of $\alpha$ to test statistical confidence (for a log-log
plot see the supplementary information Fig.~S3).
	}
	\label{fig:4}
\end{figure}

On the other hand, universal scaling can be directly tested for our tunneling
spectra. In Fig.~\ref{fig:4} the experimental $dI/dV$ data are normalized to
$T^{0.41}$ and plotted versus $eV/k_\text{B}T$. Unambiguously all spectra
collapse onto a single universal curve (red curve) as predicted for a
TLL according to Eq.~\eqref{eq:TLL_LDOS} \cite{Bockrath1999,Blumenstein2011}.
Our observation leads to the conclusion that the ZBA is a result of intrinsic
properties of the 1D edge channel and cannot arise from the DCB scenario.
Ultimately, our observations provide critical proof to render bismuthene the first
2D-TI system for which a TLL behavior has been established and 
studied on the atomic scale. Moreover, the absence of Friedel oscillations is in 
agreement with the TLL in bismuthene being helical.


Within TLL theory the strength of the Coulomb interaction $W$ between 
the electrons in the 1D edge channel can be expressed by the parameter
$K \approx \left[1+ W/\pi \hbar v_\text{F} \right]^{-1/2}$, where 
$v_\text{F}$ is the Fermi velocity. For non-interacting electrons $K=1$.
Moderate to strong electronic interactions result in two-particle scattering 
in the edge channels of the 2D-TI (see Ref.~\cite{Xu2006,Wu2006}), rendering $K$ essential for the interpretation of transport 
measurements on 2D-TIs. To date, only few experimental studies have 
addressed the effect of electronic correlations in the edge channels.
A notable exception is a recent transport study of InAs/GaSb QWs which 
reported experimental evidence for a helical TLL \cite{Li2015}.
However, the extraction of $K$ from transport data is ambiguous, 
because the resulting values vary widely depending on the underlying 
theory ($K = 0.22$~\cite{Li2015} vs.~$K = 0.8$~\cite{Vaeyrynen2016}). 
Many authors propose a more direct determination of $K$ by 
local spectroscopy (i.e. STS) of the electronic spectral function 
\cite{Eggert2000,Blumenstein2011,Bockrath1999}. 
Accordingly, for a helical TLL the characteristic power-law exponent 
of the generic spectrum \eqref{eq:TLL_LDOS} is directly related 
to $K$ via $\alpha= \frac{1}{2}(K + 1/K - 2)$ (note the prefactor 
$\frac{1}{2}$ which for a spinful TLL would be $\frac{1}{4}$ \cite{Braunecker2012}).
The experimentally determined exponent $\alpha$ thus 
translates into the helical TLL parameter $K = 0.42 \pm 0.05$ for the 
bismuthene armchair edge channels.


Qualitatively, the effective electronic interaction strength is
expected to be enhanced (reduced) with decreasing (increasing) 
Fermi velocity $v_\text{F}$ and confinement length $\xi$ of the 1D 
edge channels. For example, for HgTe/CdTe QWs ($v_\text{F} = 
5.3 \times 10^5$m/s, $\xi = 30$nm) theoretical estimates 
yield $K \approx 0.8$ (Ref.~\cite{Teo2009}), 
placing this system in the weakly to nearly non-interacting regime.
In contrast, for bismuthene the key parameters are 
$v_\text{F} = 5.5 \times 10^5 \text{m/s}$ and $\xi \approx 0.4 \text{nm}$ 
\cite{Reis2017}. Such exceptionally narrow confinement length $\xi$ 
(2 orders of magnitude less than in HgTe/CdTe QWs) is driven by the large 
bulk gap of $\Delta \sim \SI{0.8}{\electronvolt}$ in bismuthene ($\xi \approx \hbar v_F/\Delta$). 
This suggests a substantially stronger interaction, which
is nicely confirmed by our experimental value $K=0.42$. 
It is further underpinned by a theoretical model for 
bismuthene which results in $K \approx 0.5\pm 0.1$
(see supplementary information).


Bismuthene on SiC is thus the first 2D-TI candidate in which 
the spectral properties of a (helical) TLL have been verified 
in its 1D edge states. The experimentally determined 
helical TLL parameter $K$ indicates that the edge channels -- while 
topologically protected against SPB -- 
display pronounced two-particle scattering. With regard to the QSHE, bismuthene
is thus a model system to study in which way the conductance quantization is affected by these
correlation effects. \\

{\noindent
	\textbf{Methods}
}
\begin{scriptsize}
	Bismuthene was grown on n-doped 4H-SiC(0001) substrates with a resistivity of $\SI{0.01}{\ohm \cm} - \SI{0.03}{\ohm \cm}$ at room-temperature. The dopant concentration of the substrate is $\SI{5e18}{\per \cubic\centi\meter} - \SI{1e19}{\per \cubic\centi\meter}$. The STM/STS measurements have been performed with a commercial low-temperature
	STM from Scienta Omicron GmbH under UHV conditions ($p_\text{base} = \SI{2
		e-11}{\milli \bar}$). Topographic STM images are recorded as constant current
	images. After stabilizing the tip at a voltage and current set-point
	$V_\text{set}$ and $I_\text{set}$, respectively, the feedback loop is opened and
	STS spectra are obtained making additional use of a standard lock-in technique
	with a modulation frequency of \SI{787}{\hertz} and modulation amplitude
	$V_\text{mod} = \SI{1}{\milli \volt}$. Because of external modulation the
	instrumental resolution results in an gaussian energy broadening (FWHM: $\delta
	E = 2.5 e V_\text{mod}$) in addition to the thermal broadening caused by
	the Fermi-Dirac function (FWHM $=3.5k_B T$). We only show single-point spectra 
	and avoid averaging over various STS spectra of a finite sample area. 
	Prior to every measurement on bismuthene we assured that the
	tip LDOS is metallic and to a good approximation can be considered constant in
	the energy range $-\SI{100}{\milli \electronvolt} \leq E-E_\text{F} \leq
	\SI{100}{\milli \electronvolt}$ by a reference measurement on a silver surface. 
	As a result, $dI/dV(V)$ curves measured with this tip are a good measure of the 
	sample LDOS. \\
\end{scriptsize} 

{\noindent
	\textbf{Acknowledgements}
}
\begin{scriptsize}
	We thank B. Trauzettel, A. Kowalewski and J. Maciejko for useful
	discussions. This work was supported by the Deutsche Forschungsgemeinschaft
	(DFG) through the Collaborative Research Center SFB 1170
	"ToCoTronics" in W\"urzburg, the SPP 1666 Priority Program
	"Topological Insulators", and by the European Research Council (ERC) through starting
	grant ERC-StG-Thomale-336012 "Topolectrics".\\
\end{scriptsize} 

{\noindent
	\textbf{Author contributions}
}
\begin{scriptsize}
	R.S. and F.R. carried out the measurements, R.S. analysed the data and made the figures. T.M., T.H.,
	T.S, and R.T. developed the theory for the helical edge with dielectric screening.
	J.S. conceived the experiment.
	R.S. and R.C. wrote the text with input from J.S. and F.R., and all authors contributed to critical discussion of the data.
\end{scriptsize} 
\newpage
\begin{small}
	\bibliographystyle{apsrev4-1}
	\bibliographystyle{apsrev4-1}
\end{small}


\pagebreak

\onecolumngrid
\begin{center}
	\textbf{\large Supplementary Information\\
		---\\
		Tomonaga-Luttinger liquid in the edge channels of a quantum spin Hall insulator}\\[.2cm]
	R. St\"uhler,$^{1}$ F. Reis,$^{1}$ T. M\"uller,$^{2}$ T. Helbig,$^{2}$ T. Schwemmer,$^{2}$ R. Thomale,$^{2}$ J. Sch\"afer,$^{1,*}$ and R. Claessen$^{1}$\\[.1cm]
	{\itshape ${}^1$Physikalisches Institut and R\"ontgen Research Center for Complex Material Systems,\\Universit\"at W\"urzburg, D-97074 W\"urzburg, Germany\\
	${}^2$Institut f\"ur Theoretische Physik und Astrophysik, \\Universit\"at W\"urzburg, D-97074 W\"urzburg, Germany\\}
	${}^*$e-mail: joerg.schaefer@physik.uni-wuerzburg.de\\
	(Dated: \today)\\[1cm]
\end{center}
\twocolumngrid

\setcounter{equation}{0}
\setcounter{figure}{0}
\setcounter{table}{0}
\setcounter{page}{1}
\renewcommand{\theequation}{S\arabic{equation}}
\renewcommand{\thefigure}{S\arabic{figure}}
\renewcommand{\bibnumfmt}[1]{[S#1]}
\renewcommand{\citenumfont}[1]{S#1}

\onecolumngrid
\clearpage
\subsection{Structural considerations for an bismuthene armchair edge at a SiC zigzag terrace step}
\begin{figure}[H]
	\centering
	\includegraphics[width = 0.85\textwidth ]{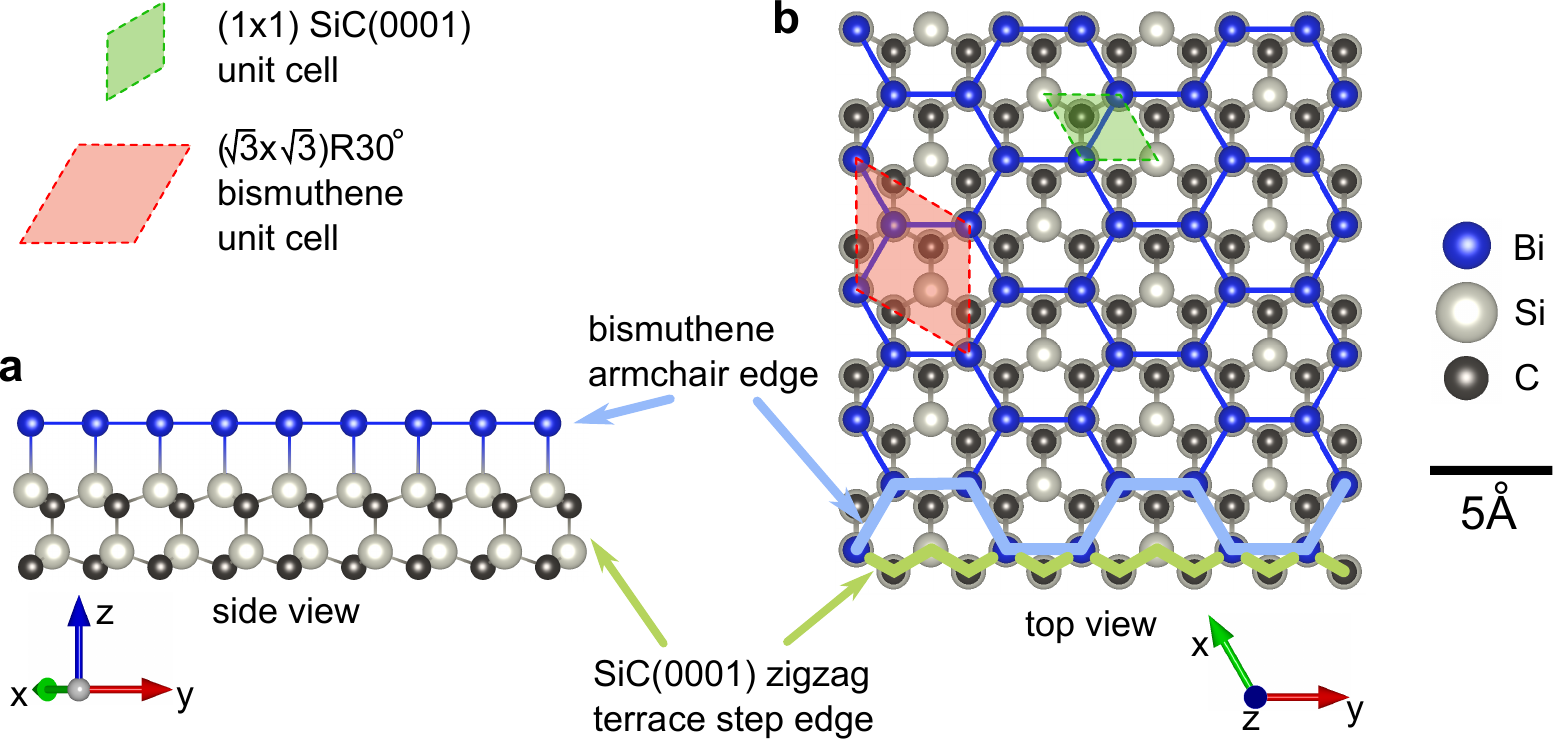}
	\caption{\textbf{Schematic overview fo bismuthene synthesized on SiC(0001)} \textbf{a}, Side view of the layered system. \textbf{b}, Top view of the geometric placement of the bismuthene layer in a commensurate $\sqrt{3}\times \sqrt{3}R30^\circ$ reconstruction. Two of three top layer Si atoms are binding to Bi. The in-plane lattice constant amounts to \SI{5.35}{\angstrom}. Due to the underlying substrate bismuthene grows in a fully planar configuration. At SiC zigzag terrace step edges the bismuthene film is naturally terminated in an armchair configuration. }
	\label{Fig:Supp5}
\end{figure}

Bismuthene is synthesized on a hydrogen-etched SiC(0001) substrate on which it forms a $\sqrt{3}\times \sqrt{3}R30^\circ$ superstructure of Bi atoms in honeycomb geometry with a lattice constant of \SI{5.35}{\angstrom}. The tensile stress of $\approx 18\%$ compared to a buckled Bi(111) monolayer causes a fully planar configuration which can be seen in the scanning tunneling microscopy (STM) constant current image by the absence of intensity corrugation between Bi atoms on honeycomb sub-lattice sites $A$ and $B$. The bismuthene film covers the entire substrate smoothly and is naturally terminated by SiC terrace steps. The SiC edges which form a zigzag configuration typically have a length of several \SI{100}{\nano \meter}. Because of bismuthene forming a $\sqrt{3}\times \sqrt{3}R30^\circ$ superstructure on SiC the experimentally rare situation of armchair edges is realized in this material system (Fig.~\ref{Fig:Supp5}). \par
\clearpage
\subsection{Estimate of the Luttinger parameter $K$} 
Helical edge states of a 2D-TI, in principle, resemble a prototypical 1D system, as their direction of motion is limited to the direction along the edge (in our case the SiC terrace step).
Topologically protected edge states decay exponentially into the bulk and are confined to the edge, where the strength of the confinement, and hence the effective width of the channel, depends on the bulk band gap and the Fermi velocity. Immediate experimental evidence for helical TLLs has so far been limited to InAs/GaSb QWs \cite{Li2015S}.
Determining the interaction parameter $K$ of the Luttinger theory, which manifests itself in observable quantities (such as velocity renormalization and  the power-law dependence of the LDOS upon approaching $E_\text{F}$) is challenging: Values on transport measurements crucially depend on the employed theoretical phenomenology, as it is the case for InAs/GaSb QWs \cite{Li2015S,Vaeyrynen2016S}.
Table \ref{tab:K parameters} provides an overview of theoretical estimates and measurements of $K$ in realizable experimental systems.
\begin{table}[H]
	\begin{ruledtabular}
		\begin{tabular}{clll}
			& HgTe/CdTe QWs   & InAs/GaSb QWs                          & bismuthene                      \\
			&                 & (Transport measurements)								& (Spectroscopic measurements)\\
			\addlinespace[0.15cm]
			\colrule
			\addlinespace[0.15cm]
			Experiment & no data         & \(K=0.21\text{ \cite{Li2015S} vs. }K=0.8\) \cite{Vaeyrynen2016S}    & \(K = 0.42\,\pm\,0.05\)     \\
			&&&\\
			Theoretical estimate    & \(K=0.8\) \cite{Teo2009S}  & \(K=0.22\text{ \cite{Li2015S}}\)                             & \(K = 0.5\,\pm\,0.1\)       \\
		\end{tabular}
	\end{ruledtabular}
	\caption{Theoretical predictions and experimental data for the \(K\)-Parameter for experimentally realizable helical TLLs.}
	\label{tab:K parameters}
\end{table}
The more immediate approach to measure $K$ from temperature scaling behavior relies on probing the electronic spectral function via STS \cite{Eggert2000S,Blumenstein2011S,Bockrath1999S}, as pursued in our work.	This requires the 2D-TI system's surface to be accessible for local tunneling measurements.
The characteristic spin-charge-separation which is inherent to spinful TLLs \cite{Calzona2015} is absent for helical TLLs because spinon and holon velocities are equal as a result of spin-momentum locking. Therefore, the characteristic fingerprint of a helical TLL is the universal scaling behaviour of the LDOS as a function of energy and temperature.
Inspired by Teo and Kane \cite{Teo2009S}, we derive an estimate of $K$ for the
one-dimensional TI channel. We start from a low energy (long wavelength) description of the system and model the interaction by a momentum independent long-ranged Coulomb interaction.

In the low-energy limit, the Luttinger Parameter is found to be~\cite{Giamarchi2003}
\begin{align}
K = \left[\frac{1+y_2-y_4}{1+y_2+y_4}\right]^{1/2} \label{eq:Kdef}
\text{ , } \; y_i = \frac{g_i}{2\pi \hbar v_F} \text{.} \tag{S1}
\end{align}
Here, $g_4$ ($g_2$) denote forward scattering amplitudes for  electrons at the same (opposite) Fermi points. For the isotropic Coulomb interaction they are equal to the long wavelength component of the potential given by
\begin{align}
g_2 = g_4 = V(q=0) = \left[\int \ddl y \ V(y) \, e^{-i q y} \right]_{q=0}\text{.} \tag{S2}
\end{align}
Elastic backscattering is prohibited via spin-momentum locking and preserved time-reversal symmetry. Furthermore, Umklapp processes are assumed to be negligible at generic incommensurate filling.

\begin{figure}[H]
	\centering
	\includegraphics{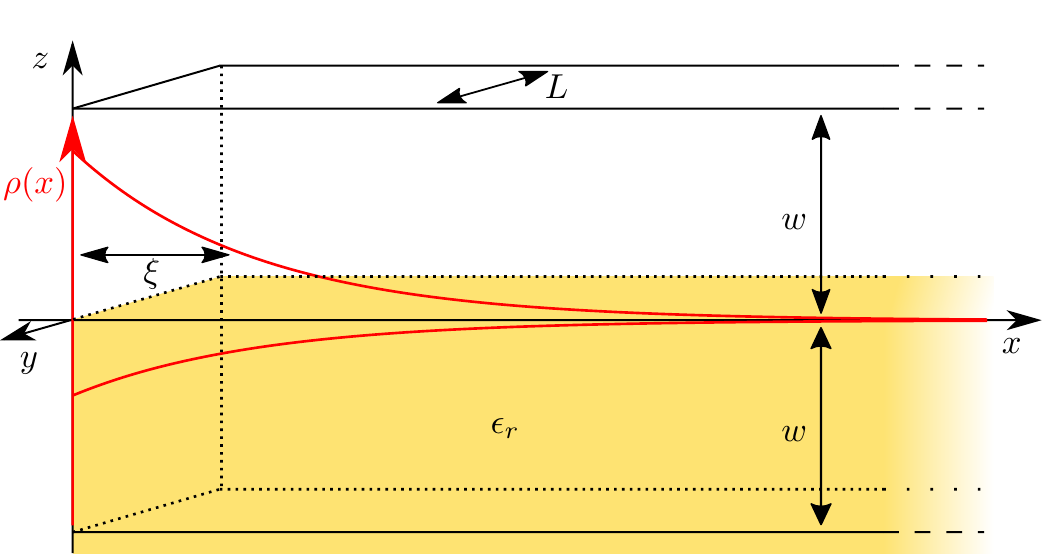}
	\caption{\textbf{Model of the charge density $\rho(x)$ of the Bismuthene edge channel on a SiC substrate (yellow) used to estimate the Luttinger parameter $K$.} The charge density $\rho(x)$ is uniformly distributed over length $L$ in $y$- and width $w$ in $z$-direction perpendicular to the interface. In $x$-direction  it is subject to an exponential decay into the bismuthene bulk with a localization length $\xi$. A mirror charge with identical dimensions, opposite sign and smaller value is induced in the substrate of relative permittivity $\epsilon_r$.  \label{fig:coulomb_setup}}
\end{figure}

We model the bismuthene edge state as a three-dimensional charge distribution (see Fig.~\ref{fig:coulomb_setup}), which be homogeneous over the edge length $L$ in $y$-direction and the width $w$ in $z$-direction perpendicular to the bismuthene surface. The charge distribution is exponentially decaying into the bulk in $x$-direction with a localization length $\xi$ and given by

\begin{align}
\rho (x,y,z) =\frac{e}{w \, \xi \,  L} \,\text{e}^{-x/\xi}\, \Theta(L/2-|y|) \, \Theta(z) \, \Theta(w-z)\text{.} \tag{S3}
\end{align}

Placing this charge distribution on top of a substrate with relative permittivity $\epsilon_r$ induces a mirror charge 
$  \rho_m (x,y,z) = -a(\epsilon_r) \, \rho (x,y,-z)$, scaled by the factor $a(\epsilon_r) = (\epsilon_r-1)/(\epsilon_r+1)$.
The electrostatic energy \(E_\text{C}\) of the system is composed of two parts.
The first component is given by the unscreened self interaction of the charge in the edge channel.
It is reduced by the second component which models the screening of the dielectric through an interaction between the channel and $\rho_m$.
The electrostatic Coulomb energy therefore amounts to
\begin{align}
E_\text{C} =  \frac{V(q=0)}{L} = \frac{1}{4 \pi \epsilon_0}   \int \ddl^3 \vect{r} \, \int \ddl^3 \vect{r}' \ \left(\frac{\rho(\vect{r})\, \rho(\vect{r'}) }{|\vect{r}-\vect{r}'|}
+ \frac{\rho(\vect{r})\, \rho_m(\vect{r'})}{|\vect{r}-\vect{r}'|} \right)\text{.} \tag{S4} \label{eq:energy}
\end{align}
As the integral in Eq.~\eqref{eq:energy} cannot be solved analytically, we will focus on two tractable limiting cases.
First, we assume $\xi \ll w \ll L$, i.\,e. a negligible decay length. This reduces the dimensionality of the problem and results in the integral expression
\begin{align*}
V_{\xi\rightarrow 0}(q=0) =
\frac{e^2}{4 \pi \epsilon_0}\frac{1}{L}
\int_0^{L} \ddl y' \int_0^{L} \ddl y \ \frac{1}{w^2} \int_0^w \ddl z' \int_0^w \ddl z \
\left(
\frac{1}{|\vect{r}_{x=0}-\vect{r}'_{x'=0}|}
- \frac{a(\epsilon_r)}{\sqrt{(y-y')^2+(z+z')^2}} \right) \text{.} \tag{S5}
\end{align*}
We can simplify the integrated expression by expanding in $w/L$ and find
\begin{equation}
V_{\xi\rightarrow 0}(q=0) =  -
\frac{e^2}{\pi \epsilon_0 (\epsilon_r+1)} \ln\left(e^{-\frac{1}{2}} \,  2^{-\epsilon_r}  \, \frac{w}{L}\right)\text{.} \tag{S6} \label{eq:limit1}
\end{equation}

Second, we assume the perpendicular width of the channel $w$ to be negligible, $w \ll \xi  \ll L$, where Eq.~\eqref{eq:energy} reduces to
\begin{align*}
V_{w\rightarrow 0}(q=0) =
\frac{e^2}{4 \pi \epsilon_0} \frac{1}{L}
\int_0^{L} \ddl y' \int_0^{L} \ddl y \ \frac{1}{\xi^2} \int_0^\infty \ddl x' \int_0^\infty \ddl x \
e^{-\frac{x+x'}{\xi}} \left(
\frac{1           }{|\vect{r}_{z=0}-\vect{r}'_{z'=0}|} - \frac{a(\epsilon_r)}{|\vect{r}_{z=0}-\vect{r}'_{z'=0}|} \right) \text{.} \tag{S7}
\end{align*}
Here, we can expand in $\xi/L$ and find
\begin{align}
V_{w\rightarrow 0}(q=0) = - \frac{e^2}{ \pi \epsilon_0 (\epsilon_r+1)} \ln\left( e^{1-\gamma} \, 2^{-1} \,
\frac{\xi}{L} \right), \tag{S8} \label{eq:limit2}
\end{align}

where $\gamma$ denotes Euler's constant.
Combining both limits from Eqs.~\eqref{eq:limit1} and \eqref{eq:limit2} through a logarithmic interpolation yields
\begin{align}
K = \left[1+\frac{V(q=0)}{\pi \, \hbar \, v_F}\right]^{-1/2} = \left[
1 - \frac{e^2 \, }{ \pi^2 \, \hbar \, v_F \, \epsilon_0 (\epsilon_r+1)}
\ln \left(
e^{-\frac{1}{2}} \,  2^{-\epsilon_r}  \, \frac{w}{L} + e^{1-\gamma} \,2^{-1}\, \frac{\xi}{L}
\right)
\right]^{-1/2}\text{.} \tag{S9}
\label{eq:formula}
\end{align}
This final result only deviates negligibly from a numerical evaluation of the full expression~\eqref{eq:energy}. A theoretical estimate of $K$ in HgTe/CdTe QWs was performed in the work by Teo and Kane~\cite{Teo2009S}, who employ screening by a metallic gate in the vicinity of the edge channel and obtain a similar logarithmic interpolation formula. Using the experimental parameters $v_\text{F} =\SI{5.3e5}{\meter \per \second}$ and $\xi = \SI{30}{\nano \meter}$, they estimate the Luttinger parameter as $K \approx \SI{0.8}{}$, which indicates rather weak interactions. In contrast, we restrict ourselves to a dielectric screening by the SiC substrate, that keeps the long range character of the Coulomb interaction intact.
The logarithmic divergence of the electrostatic energy of the edge channel in the limit of $L\rightarrow\infty$ reflects the incomplete screening of the long-ranged Coulomb interaction.
Our result is consistent with the observation that $K$ vanishes for infinitely long systems if metallic screening is neglected.~\cite{Schulz1994}
Eq. \eqref{eq:formula} is therefore only valid if $L$ is small compared to the distance between metallic gates and the edge channel.
The explicit estimate for the bismuthene edge mode is obtained using $w=\SI{0.3}{\nano \meter}$, $\xi =\SI{0.41}{\nano \meter}$, $v_F = \SI{5.5e5}{\meter \per \second}$ as experimental parameters. By assuming a range of $L=\SIrange{e-8}{e-6}{\meter}$ for the edge channel length and $\epsilon_r=\si{10}$ for the relative permittivity of the SiC substrate \cite{Patrick1970}, we obtain a Luttinger parameter of \(K= 0.5 \pm 0.1\).
This places the 1D edge channel of bismuthene beyond the regime of weak interactions. It should be noted that $\epsilon_r=\si{10}$ for undoped SiC is an estimate in this simple modeling, as $\epsilon_r$ is known to increase for heavily doped semiconductors \cite{Dhar1985}, which would result in enhanced screening. Notwithstanding further possible refinements of the current theoretical analysis, our approximation puts the theoretical prediction within the range of the Luttinger parameter extracted from STS scaling in experiment (see Tab. \ref{tab:K parameters}). Overall, the central parameter that governs the strong decrease in K from HgTe/CdTe QWs to bismuthene is the reduced channel width $\xi$, which implies stronger Coulomb interactions induced by a substantially increased confinement of the quasi-1D DOS of the helical edge mode.

\clearpage
\subsection{STS spectra in close-up analysis}
\subsubsection*{a) Logarithmic plot}
\begin{figure}[H]
	\centering
	\includegraphics[width = 0.4\textwidth ]{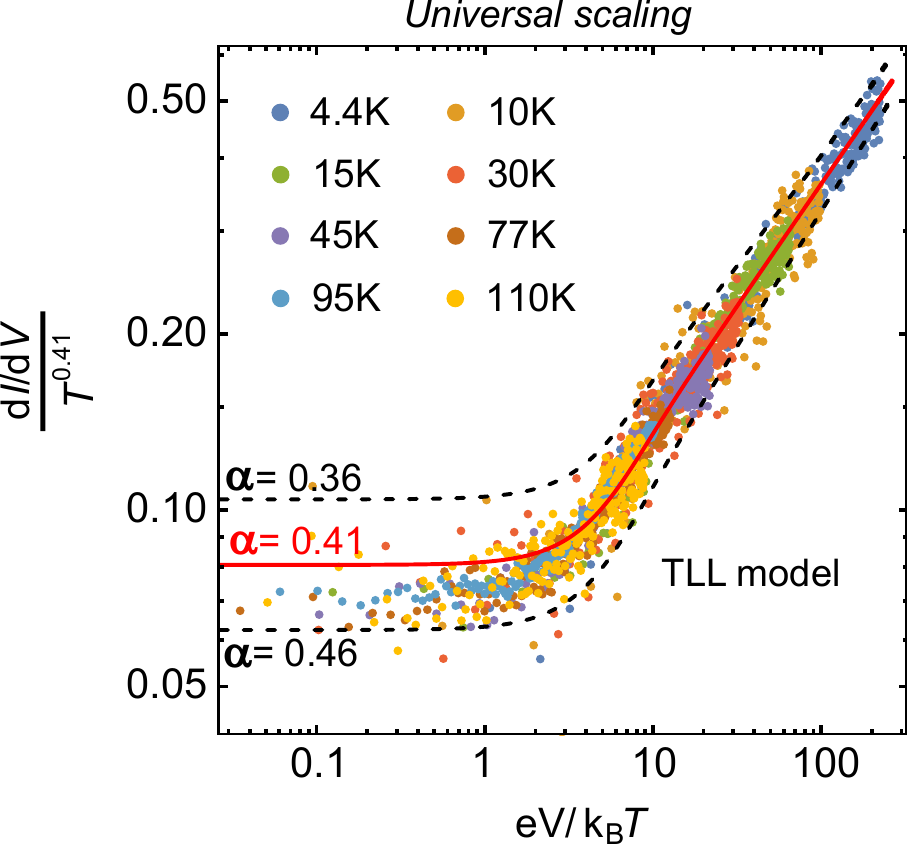}
	\caption{\textbf{Universal scaling of the ZBA: hallmark of a TLL.} Double-logarithmic plot of the rescaled ZBA spectra of Fig.~4. All spectra collapse onto a single universal curve predicted as a hallmark for tunneling into a TLL. Red curve: $\rho_{\text{TLL}}$ according to
		Eq.~(1) (taking into account thermal and instrumental broadening) with exponent $\alpha = 0.41$. Black dashed curve: for $\pm 0.05$ variation of $\alpha$ to test statistical confidence.}
	\label{Fig:Supp3}
\end{figure}
\subsubsection*{b) Set-point variation}
\begin{figure}[H]
	\centering
	\includegraphics[width = 0.8\textwidth ]{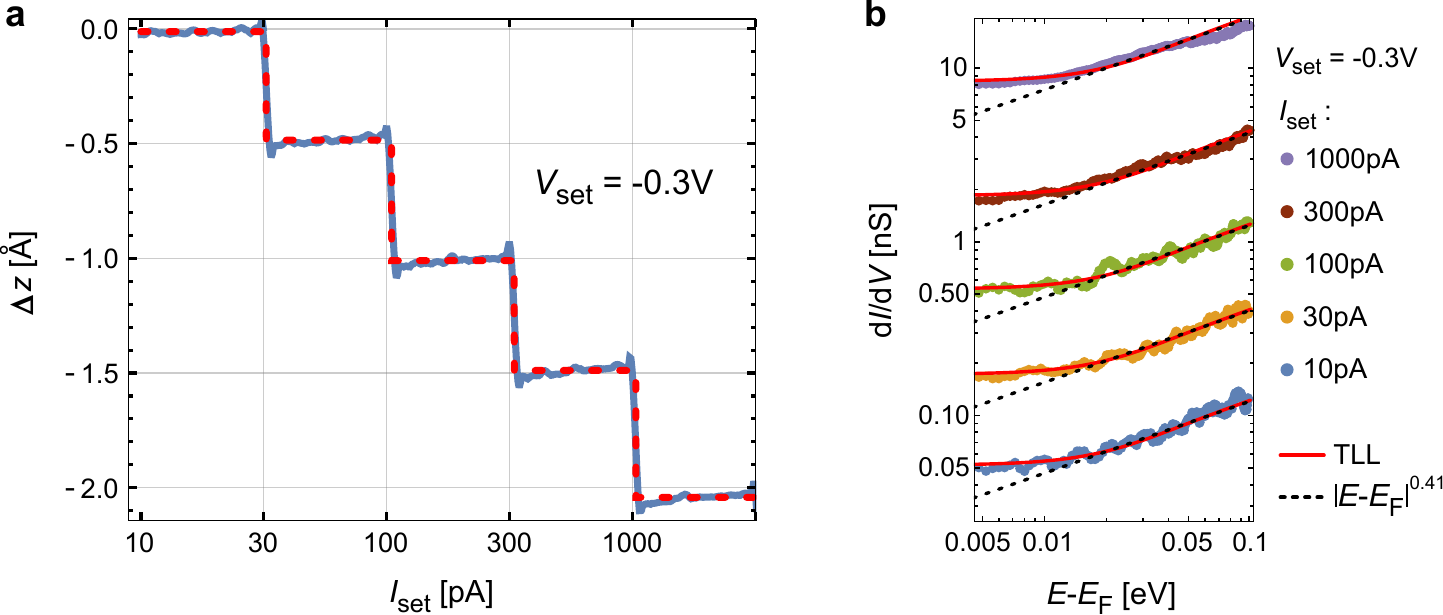}
	\caption{\textbf{Influence of $I_{\text{set}}$ on the power-law of the ZBA.} \textbf{a}, Tip-to-sample variation $\Delta z$ as a function of the set-point current $I_{\text{set}}$ at fixed set-point bias voltage $V_{\text{set}} = \SI{-0.3}{\volt}$ and at a fixed bismuthene edge position. This nicely reflects that the tunneling current depends exponentially on the tip-to-sample distance. \textbf{b}, Double-logarithmic plot of the ZBA at $T = \SI{77}{\kelvin}$ measured for different set-point currents $I_{\text{set}}$ and therefore also the tip-to-sample distance according to a. Black dashed line: power-law dependence $dI/dV \sim |E-E_\text{F}|^{0.41}$. Red line: TLL model according to Eq.~(1) (taking into account thermal and instrumental broadening).}
	\label{Fig:Supp4}
\end{figure}

It is known that the electric field of the tunneling tip can induce band bending effects if the tunneling current is too high and the tip-to-sample distance $z$ becomes to small. We excluded possible tip induced effects on the power-law scaling of the ZBA by measuring spectra for different set-point voltages and currents ranging from $-\SI{0.1}{\volt} \leq V_{\text{set}} \leq -\SI{0.3}{\volt}$ and $\SI{0.01}{\nano \ampere} \leq I_{\text{set}} \leq \SI{1.0}{\nano \ampere}$ which apart from an overall scaling factor had no significant influence, see Fig.~\ref{Fig:Supp4}. 
\clearpage
\subsection{Alternative Models for a Zero Bias Anomaly}
The characteristic feature of a TLL is the power-law decay of the LDOS at low-energies. Moreover, $dI/dV(V)$ spectra for different temperatures collapse onto an universal scaling curve as in Fig.~4 and \ref{Fig:Supp3}. Yet, there are also alternative mechanisms that could potentially lead to a ZBA.\par 
\subsubsection*{a) Efros-Schklovskii like Coulomb Pseudogap}
In a low-dimensional metallic system the presence of disorder can lead to a suppression of spectral weight around zero energy. The phenomenon is an example for an Efros-Shklovskii like Coulomb pseudogap. The expected LDOS can be found to exhibit an exponential suppression according to~\cite{Bartosch2002S}
\begin{align}\label{eq:Bartosch51}
\rho(V, T) =& \rho_0 \coth\left(\frac{e V}{2 k_B T}\right) 2 T \int_{0}^{\infty} \frac{\sin(e V t/k_B)\cos(\sqrt{2 e \Omega t /k_B)}}{\sinh(\pi t T)} \times \notag \\ 
&\times \exp\left[ - \sqrt{\frac{\Omega}{\pi}} \int_{0}^{\infty} \frac{1-\cos(V' t )}{V'^{3/2} \tanh(e V'/2 k_B T)}dV' \right] dt,\tag{S11}
\end{align} 
where $\Omega := f_0^2/32 \pi D_0$ depends on intrinsic properties of the system only, i.e., the electron-electron interaction $f_0$ and the diffusion coefficient $D_0$, which reflects the strength of disorder in the system. The parameter $\Omega$ will serve as a fitting parameter to our experimental data. \par
\begin{figure}[H]
	\centering
	\includegraphics[width = 0.4\textwidth ]{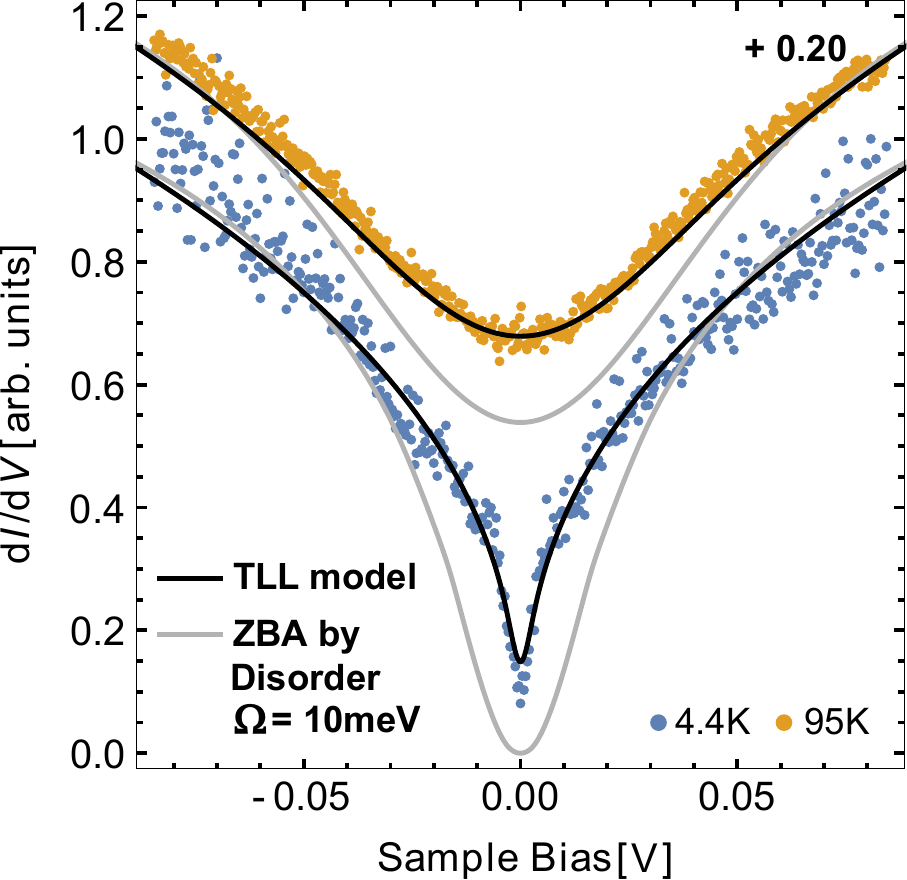}
	\caption{\textbf{Comparison with the disorder induced ZBA model.} The measured ZBA in bismuthene at $\SI{4.36}{\kelvin}$ (orange dots) and at $\SI{95}{\kelvin}$ (blue dots) is compared with the expected $dI/dV(V)$ spectrum from the TLL model (dashed black line) and the model for a disorder induced ZBA model (gray line). The spectra were normalized at $V = \SI{100}{\milli \electronvolt}$ and the spectrum taken at $\SI{95}{\kelvin}$ is shifted by $0.20$. In the case of the disorder induced ZBA close matching can not be achieved over the whole energy range. A coarse approximation to the data can be achieved at higher binding energies, but the match is increasingly poor at lower energies around zero-bias. On the other hand, the TLL model reproduces the spectrum over the entire energy range with $\alpha = 0.41$. }
	\label{Fig:Supp1}
\end{figure}

In contrast to the TLL model, however, the functional form of the arising ZBA in this scenario is described by an exponential decrease of the $dI/dV$ for low-energies. The functional form of the Coulomb pseudogap model according to Eq.~\eqref{eq:Bartosch51} with the interaction strength $\Omega$ as the only free parameter is shown in Fig.~\ref{Fig:Supp1}. Whereas a coarse approximation to the data can be achieved for higher sample bias, only poor accordance is given for lower sample bias around $V = 0$. Particularly, matching of the exponential form according to Eq.~\eqref{eq:Bartosch51} with the data can not be achieved over the entire bias range of the ZBA. This evidences that a disorder induced Efros-Schklovskii-like Coulomb pseudogap cannot explain the observed spectral behavior.\par
\subsubsection*{b) Dynamical Coulomb Blockade}
In case that the applied bias voltage $V$ is smaller than the resulting charging energy of the system, i.e., $ eV < E_C = e^2/C_\Sigma$ where $C_\Sigma$ is the total capacitance (described in the following), charging effects can in principle induce a ZBA in STS measurements. As an important example, one may consider an electrical circuit that contains an electromagnetic environment (e.g., the substrate) modeled by external impedances in the tunneling circuit. This type of models are usually referred to as dynamical Coulomb blockade (DCB) models.\par
\begin{figure}[H]
	\centering
	\includegraphics[width = 0.6\textwidth]{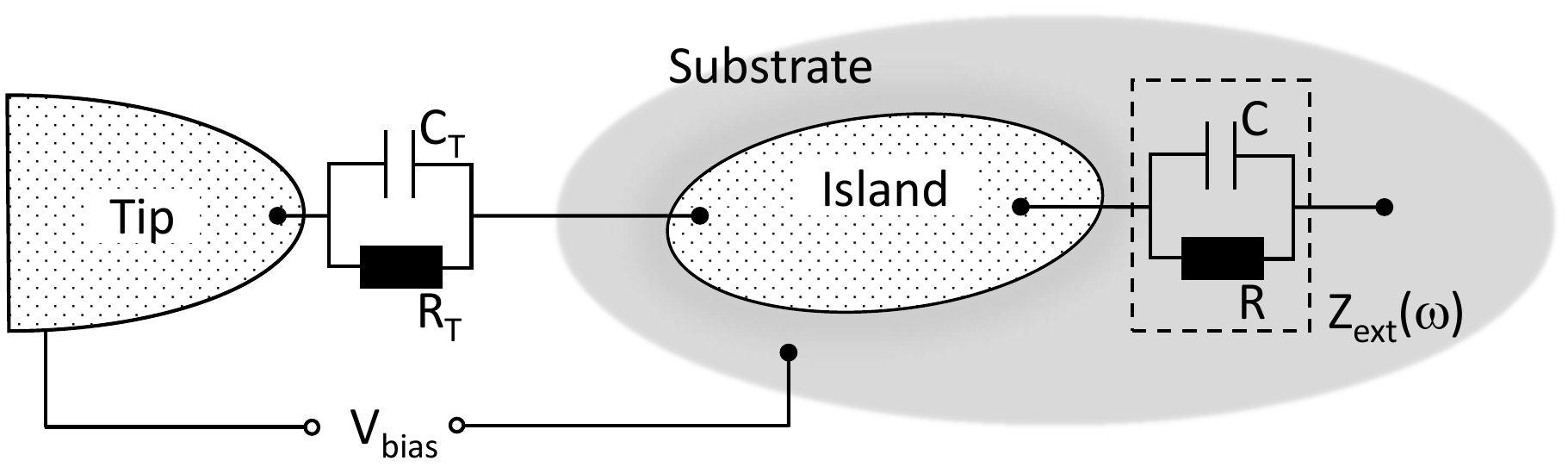}
	\caption{\textbf{Model for dynamical Coulomb blockade} Schematic picture of electron tunneling between a metallic source and drain through an intermediate island coupled to a dissipative environment $Z_{\text{ext}}(\omega)$. Tunneling can be blocked if the electrostatic energy $E_C$ of a single excess electron on the island is large compared to the energy scale of thermal fluctuations.
	}
	\label{fig:CoulombBlockade_SnAl}
\end{figure}

 
In the framework of the DCB scenario, the STS tunneling experiment can conveniently be regarded as a double junction as in Fig.~\ref{fig:CoulombBlockade_SnAl}~\cite{Brun2012, Ming2018S}. Electrons tunnel from a metallic tip onto an 'island' on the sample surface and subsequently into the substrate. Part of this double junction is a first tunneling junction (T-junction) which models the tunneling between tip and island with an effective capacitance $C_T$ in parallel with a resistance $R_T$. The tunneling resistance $R_T$ can be inferred from the set-point current and voltage as $R_T = V_{\text{set}}/I_{\text{set}}$, and is typically $ \SI{0.1}{\giga \ohm} \lesssim R_T \lesssim \SI{10}{\giga \ohm}$ in our experiments. The tunneling capacitance $C_T$ is usually not known, but typically on the order of $C_T \lesssim \SI{1}{\atto \farad}$.~\cite{Brun2012, Ming2018S} The second junction which models a tunneling process of electrons from the sample surface into the substrate is a second $RC$-junction (S-junction) in series with the T-junction. The resistance $R$ and capacitance $C$ of the S-junction are usually unknown and form fit parameters in the DCB model.
The following model in terms of the DCB scenario is valid, if in an experimental STS measurement the complete bias voltage drops over the T-junction, i.e., $R_T \gg R$, and as long as $R$ is comparable to the resistance quantum $R_K = \SI{25.8}{\kilo \ohm}$. Then tunneling in the DCB regime along with an excitation of the dissipative environment $Z_{\text{ext}}(\omega)$ can be treated quantum mechanically by the so-called environment-quantum fluctuation or $P(E)$-theory~\cite{Devoret1990S}, within which the tip to sample (sample to tip) tunneling probability $\Gamma^+$ ($\Gamma^-$) at temperature $T$ can be calculated as
\begin{subequations}
	\label{eq:tunneling_prop}
	\begin{align}
	\Gamma^+ &= \frac{1}{e^2 R_T} \int_{-\infty}^{\infty} \int_{-\infty}^{\infty} \rho_t(E) \rho_s(E'-eV) f(E,T)\left[ 1-f(E' -eV,T)\right] P(E-E') dEdE'\tag{S11a}\\
	\Gamma^- &= \frac{1}{e^2 R_T} \int_{-\infty}^{\infty} \int_{-\infty}^{\infty} \rho_t(E) \rho_s(E'-eV) \left[ 1-f(E,T)\right] f(E'-eV,T) P(E'-E) dEdE',\tag{S11b}
	\end{align}
\end{subequations}
where $V$ is the bias voltage, $f$ is the Fermi function and $E$ ($E'$) are the energy levels at the tip (sample), respectively. $ \rho_t$ and $\rho_s $ are the tip and sample LDOS, respectively. $P(E-E')$ describes the probability that the electron looses the energy $E-E'$ to the dissipative environment during the tunneling process. In the case of elastic tunneling where energy is conserved within the double junction during the tunneling event the probability factor $P(E-E')$ reduces to a delta function $P(E-E') = \delta(E-E')$.\par
In principle, the $P(E)$ function can be calculated as \cite{Devoret1990S}:
\begin{equation}
P(E) = \frac{1}{2 \pi \hbar} \int_{-\infty}^{\infty} \exp\left[ J(t) + i Et/\hbar\right] dt \tag{S12}
\label{eq:P(E)}
\end{equation}
with
\begin{equation}
J(t) = 2 \int_{0}^{\infty} \frac{\operatorname{Re}\left[ Z(\omega)\right] }{\omega R_K } \frac{e^{-i\omega t} - 1}{1-e^{-\hbar\omega/k_BT}}d\omega, \tag{S13}
\label{eq:J(t)}
\end{equation}
where $ Z(\omega) = \left[ i \omega C_T + Z^{-1}_{\text{ext}}(\omega) \right]^{-1}$ is the frequency dependent complex impedance as seen from the tunnel junction~\cite{Brun2012}. The frequency dependent complex impedance of the dissipative environment $ Z_{\text{ext}}(\omega) = \left[ i \omega C + R^{-1} \right]^{-1}$, on the other hand, depends solely on $R$ and $C$ of the S-junction. Consequently, the total impedance can be written as $
Z(\omega) = \left[ i \omega C_\Sigma + R^{-1} \right]^{-1}$, with $C_\Sigma = C_T + C$. The total tunneling current is calculated as the sum of the tunneling current in both directions $ I(V) = -e \left( \Gamma^+ - \Gamma^-\right)$.\par
In the following calculations, $\rho_t$ and $\rho_s$ are both taken to be constant in the bias range $\SI{-100}{\milli \volt} < V < \SI{100}{\milli \volt}$, which is reasonable since the tip has been ensured to be metallic and it has been seen that the bismuthene edge LDOS inside the bulk gap, despite the ZBA, is approximately constant. The functional form of the differential conductivity can be calculated as:
\begin{equation}
\frac{dI}{dV}(V) = \frac{1}{R_T} \left[ 1 + 2 \int_{0}^{\infty} \frac{\pi k_B^2 T^2}{\hbar^2} \operatorname{Im}\left[ e^{J(t)} \right] \cos\left( \frac{eV t}{\hbar} \right) \text{csch}^2 \left( \frac{\pi k_B T t}{\hbar} \right)t \; dt\right]. \tag{S14}
\label{eq:dI/dV_dCb}
\end{equation}
In the case of $T\rightarrow 0$ and $V \rightarrow 0$ the $I(V)$-characteristic and the differential conductance $dI/dV(V)$ become \cite{Devoret1990S}
\begin{subequations}
	\begin{align}
	I(V) &= \frac{\exp(-2\gamma/g)}{\Gamma(2+2/g)} \frac{V}{R_T} \left( \frac{\pi e |V|}{g E_C}\right)^{2/g}\tag{S15a}\\ 
	&\rightarrow \frac{dI}{dV}(V) \sim |V|^{2/g}. \tag{S15b}
	\label{eq:DCB_powerLaw}
	\end{align}
\end{subequations}
It is apparent that in this limit the $dI/dV(V)$ is also characterized by a power-law (with a power-law exponent $2/g= 2R/R_K$), as is the case for tunneling into a TLL. This fact was pointed out by Safi \emph{et al.}~\cite{Safi2004}. 
\par
\begin{figure}[H]
	\centering
	\includegraphics[width = \textwidth]{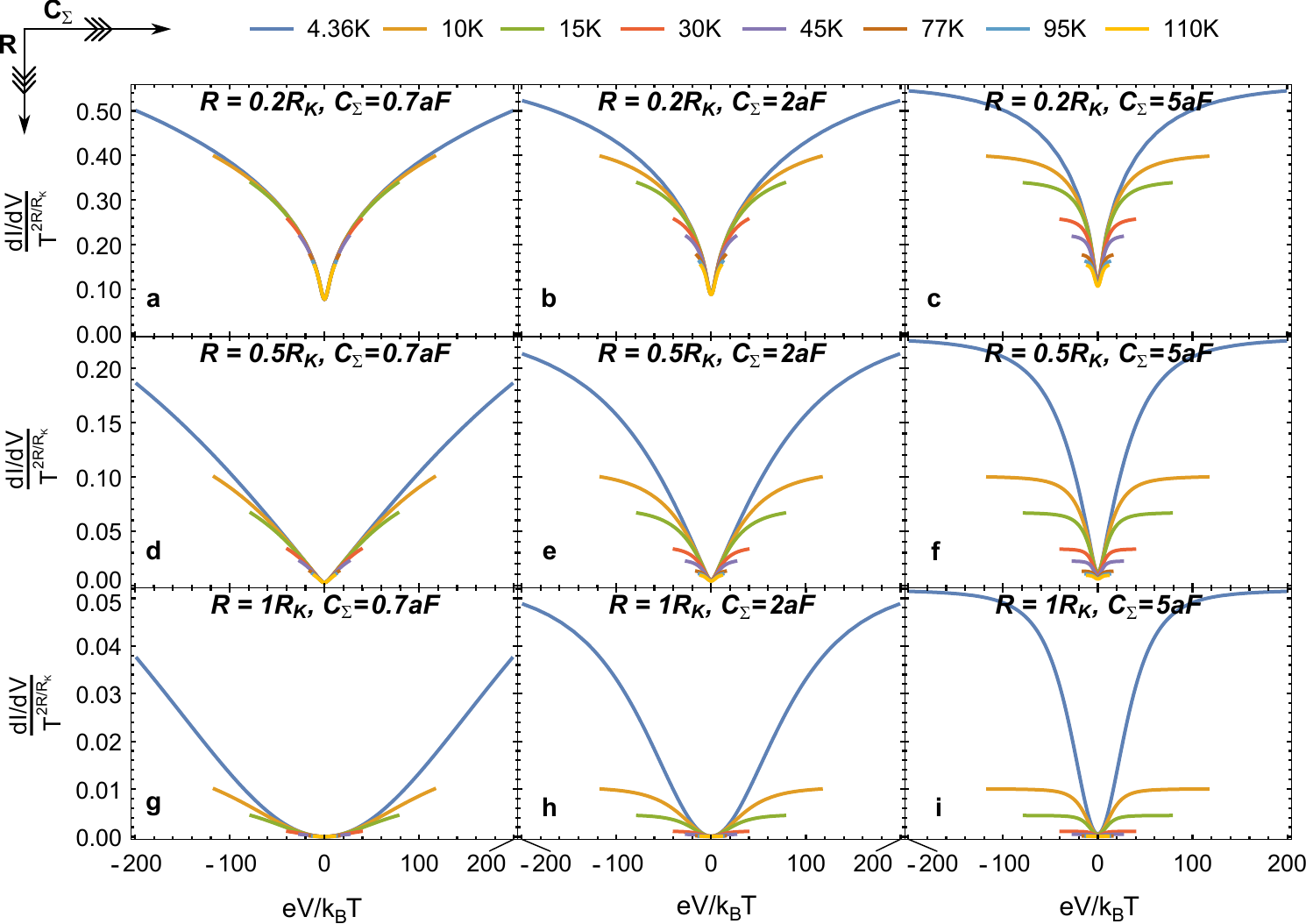}
	\caption{\textbf{Scaling of the ZBA in the DCB model.} Numerical solution of Eq.~\eqref{eq:dI/dV_dCb} for different values of $R$ and $C_\Sigma = C_T + C$ plotted on a rescaled conductivity (according to $(dI/dV)T^{-2R/R_K}$) and energy scale (according to $eV(k_BT)^{-1}$). In general, whereas $C_\Sigma$ mainly determines the depth and width of the ZBA, the main effect of $R$ is to determine the steepness of the ZBA. In the energy range $\SI{-100}{\milli \volt} < V < \SI{100}{\milli \volt}$ where we observe the ZBA in our measurements a coincidentally occurring universal scaling law, as in the case of a TLL, is only observed for $R = 0.2R_K$ and $C_\Sigma = \SI{0.7}{\atto \farad}$, where all ZBAs collapse onto a universal scaling curve for temperatures $\SI{4.4}{\kelvin} \leq T \leq \SI{110}{\kelvin}$.}
	\label{fig:dCB}
\end{figure}

For arbitrary temperatures, however, it is not possible to give an analytic solution of the $dI/dV(V)$. One has to calculate it numerically by solving Eq.~\eqref{eq:dI/dV_dCb}. In Fig.~\ref{fig:dCB} we present the result for different values of $R$ and $C_\Sigma$ and plot it on a rescaled conductivity $(dI/dV)\text{T}^{\text{-2R/R}_\text{K}}$ and energy scale $eV(k_B T)^{-1}$ in order to detect coincidental universal scaling behavior in the bias voltage range $\SI{-100}{\milli \volt} < V < \SI{100}{\milli \volt}$ (the voltage range where we observed the ZBA in bismuthene measurements). In the DCB model all ZBAs collapse onto a universal scaling curve only for parameters $R = 0.2R_K$ and $C_\Sigma = \SI{0.7}{\atto \farad}$ and for spectra in the temperature range $\SI{4}{\kelvin} < T < \SI{110}{\kelvin}$. In general, whereas $C_\Sigma = C_T + C$ mainly determines the depth and width of the ZBA, the main effect of $R$ is to determine the steepness of the ZBA.\par
We now want to discuss implications for our ZBA measurements that follow from the DCB model. First, we want to note that within the DCB model the metallic bismuthene edge channels form the capacitive island and are thought to have poor electrical contact to the SiC substrate modeled by the S-junction. A crude estimate is that bismuthene edge channels which cover a smaller area $A$ on the SiC substrate will affect the charging of the S-junction and consequently enlarge $R$ and reduce $C$ according to: $C \propto A = d\times l$ and $ R \propto A^{-1} = (d\times l)^{-1}$ for a straight metallic edge with length $l$ and width $d$ (typically $d \lesssim \SI{1.5}{\nano \meter}$, see Fig.~1(c)). Therefore, $C$ and $R$ are \emph{local} properties of the system \cite{Ming2018S}. \textit{Local} deviations should therefore manifest in \textit{local} differences of the power-law exponent $2/g$ and more importantly impede universal scaling. We measured the ZBA on four different samples and on various positions on each sample. The observation that we did not measure any significant deviations of the power-law exponent, not even for local variations of $E_\text{F}$ ($ \gtrsim \SI{100}{\milli \electronvolt}$), the size of the bismuthene domain nor variations of the set-point parameters $I_{\text{set}}$ and $V_{\text{set}}$ (Fig.~\ref{Fig:Supp4}), together with universal scaling (Fig.~4 and \ref{Fig:Supp3}), strongly points towards the ZBA being a result of intrinsic properties of the 1D edge channel and hence the exclusion of the DCB scenario. \par
\clearpage

\bibliographystyle{apsrev4-1}
\end{document}